# Experimental study on compression property of regolith analogues

Tomomi Omura and Akiko M. Nakamura

Department of Planetology, Graduate School of Science, Kobe University,
1-1 Rokkodai, Nada-ku, Kobe
657-8501, Japan

**ABSTRACT**

The compression property of regolith reflects the strength and porosity of the regolith layer on small bodies and their variations in the layer that largely influence the collisional and thermal evolution of the bodies. We conducted compression experiments and investigated the relationship between the porosity and the compression using fluffy granular samples. We focused on a low-pressure and high-porosity regime. We used tens of μm-sized irregular and spherical powders as analogs of porous regolith. The initial porosity of the samples ranged from 0.80 to 0.53. The uniaxial pressure applied to the samples lays in the range from 30 to $4\times10^5$ Pa. The porosity of the samples remained at their initial values below a threshold pressure and then decreased when the pressure exceeded the threshold. We defined this uniaxial pressure at the threshold as "yield strength". The yield strength increased as the initial porosity of a sample decreased. The yield strengths of samples consisting of irregular particles did not significantly depend on their size distributions when the samples had the same initial porosity. We compared the results of our experiments with a previously proposed theoretical model. We calculated the average interparticle force acting on contact points of constituent particles under the uniaxial pressure of yield strength using the theoretical model and compared it with theoretically estimated forces required to roll or slide the particles. The calculated interparticle force was larger than the rolling friction force and smaller than the sliding friction force. The yield strength of regolith may be constrained by these forces. Our results may be useful for planetary scientists to estimate the depth above which the porosity of a regolith layer is almost equal to that of the regolith surface and to interpret the compression property of an asteroid surface obtained by a lander.

1. INTRODUCTION

Small airless bodies are remnants of the solar system formation, while they might have metamorphosed by collisional processes in their evolution. The surface of small bodies such as asteroids is covered by regolith consisting of porous and compressive granular matter (Sears, 2015). The relationship between porosity and compression, or strength, of regolith is a key to better understand the collisional evolution of the bodies. For example, the shape of impact craters on simulated regolith is dependent upon the porosity of the regolith, probably because a projectile penetrates more deeply into the regolith target as observed for highly porous solid targets such as sintered hollow glass beads (Wada and Nakamura, 2012; Okamoto and Nakamura, 2017). At an impact velocity lower than 1 km/s, it is experimentally shown that a fraction of a rocky impactor onto regolith can survive as a block (Nagaoka et al., 2014). Impact onto regolith at the surface of a small body occurs not only with projectiles from interplanetary space but also with blocks from the body itself at low collision velocities. Intrusion depths of the blocks ejected by impact and re-accumulated onto the regolith depend on the compression property or mobility of the regolith. Such an intrusion depth may affect the collisional lifetime of blocks (Durda et al., 2011). According to a previous laboratory study, if the velocity of an impacting block onto regolith is so low that the pressure due to the impact is smaller than the compressive strength of the regolith, then the block would rebound, otherwise it would penetrate into the regolith (Machii et al., 2013; Nakamura et al., 2013). Moreover, porosity structure of regolith is crucial for thermal evolution of small bodies (Akiridge et al., 1998). The thermal conductivity of regolith is dependent not only on the grain size but also on the porosity of the regolith (Gundlach and Blum, 2013; Sakatani et al., 2016).

The porosity of regolith is directly measured for the moon, whereas it is estimated using the density obtained by radar observations for asteroids and Martian moons (Mitchell et al., 1974; Magri et al., 2001; Ostro et al., 2004; Busch et al., 2007). However, such measurements can only probe the near-surface density of regolith. It is known that core tube samples obtained from deep layers of lunar regolith have larger densities because of soil pressure and vibration of impact-induced shaking (Mitchell et al., 1974). Similar physical processes and thus the depth dependence of regolith density can be expected on asteroids. The porosity estimated for asteroidal surfaces is larger than random close packing of monodisperse particles (~0.36), namely, the surfaces are still porous enough to be further compressed (e.g., Scott and Kilgour, 1969).

The density gradient of regolith is determined by the compression properties of the regolith and the compression properties of a granular layer like regolith depend on the structure of the layer and the physical properties of the constituent granular particles. A granular layer is compacted by rearrangement, elastic and plastic deformation, and fragmentation of constituent particles. In case of compaction under a low-pressure regime, deformation and fragmentation of the constituent particles are negligible and therefore the macroscopic structure of the granular layer and forces required to rearrange the particles against interparticle forces determine the compression properties of the layer. The macroscopic structure of a granular layer, in other words, how particles are aligned in the layer, depends on the initial porosity of the layer. The initial porosity is determined by a balance between the gravitational force and the interparticle force acting on the constituent particles (Kiuchi and Nakamura, 2014). The initially loose structure is then changed against interparticle forces acting on the particles by applied compression or vibration.

Interparticle forces acting on particles depend on the following factors.

(a) The particle diameter. The particle diameter of lunar regolith was directly measured and the median diameter is 60–80 μm (Heiken et al., 1991). The particle diameters on asteroids were suggested to be tens to hundreds of microns based on polarimetric observations of asteroids (Dollfus and Zellner, 1979). Images taken by Hayabusa showed that smooth terrain consists of millimeter to centimeter sized particles (Yano et al., 2006). Gundlach and Blum (2013) estimated "typical" diameters based on thermal inertia of the asteroid and found to vary within a range from ~10 μm to cm. The particle diameters of regolith vary from one body to another and is shown to have a negative correlation with the radius of the body or its escape velocity.

Theoretically, interparticle forces (e.g. pull-off force) increase with particle diameter if the particle is a perfect sphere and the gravitational acceleration is of the order of $10^{-3}$–$10^{-2}$ m/s$^2$ on small bodies of ~10–100 km in diameter (Johnson et al., 1971). A fluffy regolith layer is, therefore, possibly formed under this low gravitational conditions even if the diameter of regolith particles is of the order of millimeters.

(b) The particle shape. The shape of most of regolith particles is expected to be irregular because regolith is formed either by accretion of impact fragments or in-situ thermal fatigue of blocks on small bodies (Delbo et al., 2014). Regolith samples returned from the moon and from the asteroid Itokawa confirmed their irregular shapes (Mitchell et al., 1974; Tsuchiyama et al., 2011).

(c) The material composition. The material composition of asteroid surfaces is mainly silicates with some exceptions. S-and C-type asteroids are considered as parent bodies of ordinary chondrites and carbonaceous chondrites, respectively, which are composed of silicates (Burbine et al., 2002).

(d) The surface chemistry. Interparticle forces of silica particles are reduced by adsorbed molecules on the particles and enhanced under ultra-high vacuum conditions (Perko et al., 2001; Kimura et al., 2015).

It is difficult to measure the compression properties of regolith layer analogues that simultaneously mimic both the particle size and the porosity because we cannot reproduce a porous structure with millimeter sized particles under earth's gravity. We can reproduce a porous structure only with smaller particles.

Compression experiments of granular materials with a relatively high porosity were conducted by several researchers. Yasui and Arakawa (2009) obtained a compression curve of a 1 μm silica powder. The initial porosity of their sample was 0.64. In our previous work, compression curves of silica and alumina powders with different particle sizes were obtained (Omura et al., 2016). The initial porosity of the samples ranged from ~0.5 to ~0.8. However, the uniaxial pressure applied to the samples in these studies was higher than ~$10^4$ Pa that exceeds the pressure range relevant to a shallow depth of a regolith layer. Compression curves at low pressure showed a regime where the samples kept their porosity almost constant (Blum et al., 2006; Guettler et al., 2009; Machii et al., 2013). The initial porosity was higher than ~0.75, but it covered only higher part of possible porosity for asteroid regolith, which is expected to be 0.4–0.9 (Kiuchi and Nakamura, 2014). Güttler et al. (2009) proposed an empirical formula to reproduce the compression curve of uniaxial compression. The formula is described by four parameters obtained by laboratory experiments; an initial filling factor under low pressure, an equilibrium filling factor under high pressure, turnover pressure, and a logarithmic width of transition from initial to equilibrium filling factors. However, the initial porosity, size distribution, shape and material composition of constituent particles, and surface conditions of regolith on an asteroid surface differ from those in the laboratory. Therefore, to estimate the compressive strength of a regolith layer on an asteroid, the effects of initial porosity, particle properties, and surface chemistry on the compression curve must be separately evaluated.

To obtain the data on the compressive properties, especially at a low-pressure and high-porosity regime, we conducted compression experiments and investigated the relationship between the porosity and the pressure using particles of different size distribution, shape, and composition with various initial porosity. We determined the threshold pressure below which the porosity of samples remains almost unchanged. We compared our results with a previously proposed theoretical model for granular material and obtained a possible constraint on the threshold pressure beyond which regolith starts to be compacted. A possible application to an estimate of porosity structure in asteroid surfaces is briefly discussed.

## 2. EXPERIMENTS

### 2.1 Samples

We used fine powders to produce fluffy samples that mimic a regolith structure with a relatively high porosity on small bodies. The selection of powders aimed at investigating the effects of particle size distribution, shape, and material composition on the compression properties of the powders. We used irregular alumina particles of three different sizes and spherical silica particles of three different sizes as our sample powders. Figure 1 shows scanning electron microscope (SEM) images of each sample powder. The physical properties of the powders are listed in Table 1. Figure 1 also presents a SEM image of fine components of fragments produced by an impact experiment in which a cylindrical iron projectile was shot at a block of basalt. The shapes of "irregular" samples resemble the shape of impact fragments, namely, an important component of regolith. Figure 2 shows the particle size distributions of our sample powders in the form of cumulative volume fractions versus particle diameter, which were determined by a laser diffractometer (SHIMAZU SALD-3000S). The sample powders have different ranges of particle size distribution. We use the median diameter as a representative particle diameter of the sample powder and the ratio $d_{85}/d_{15}$ as the width of a particle size distribution where $d_{85}$ and $d_{15}$ indicate the diameters at the cumulative volume fraction of 85% and 15%, respectively.

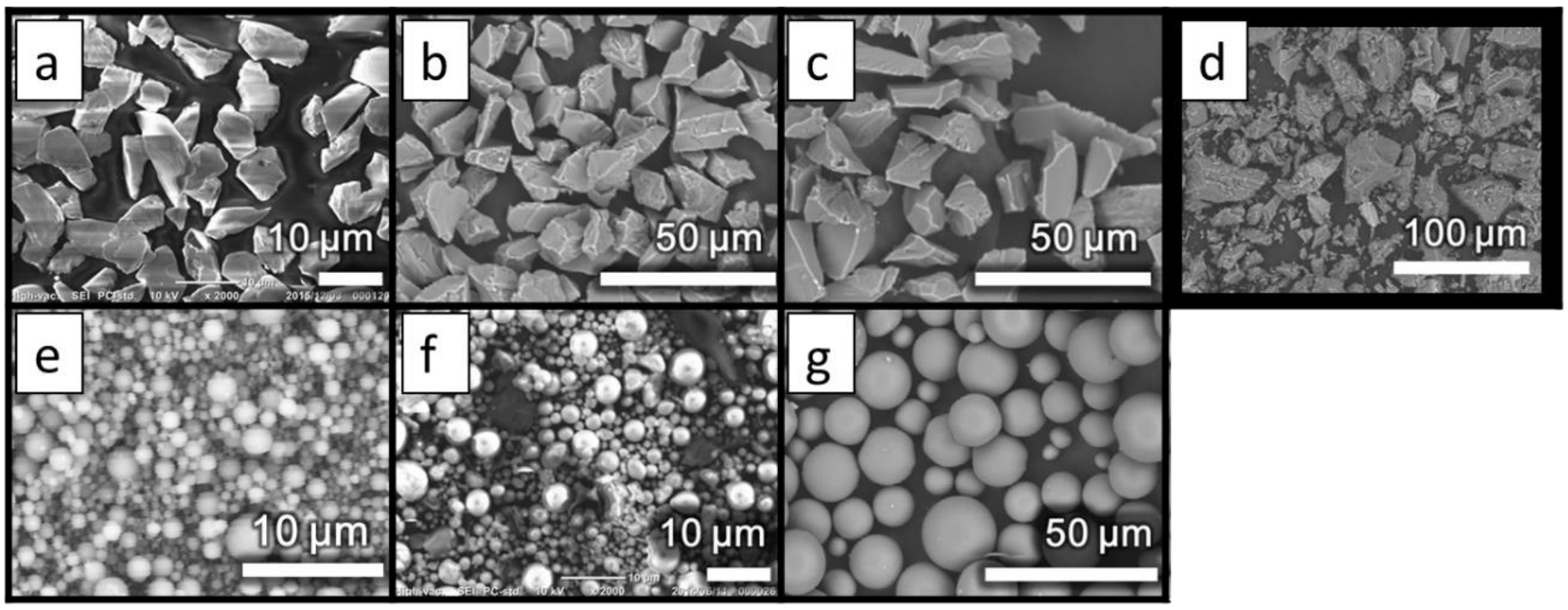

Fig. 1 SEM image of (a) alumina (6.5 µm), (b) alumina (15 µm), (c) alumina (23 µm), (d) impact fragments, (e) silica beads (1.7 µm), (f) fly ash (4.8 µm), (g) glass beads (18 µm).

Table 1 Properties of sample powder.

| Name | Particle density ($g/cm^3$) | Shape | Material | Median diameter (µm) | $d_{85}/d_{15}$ |
|---|---|---|---|---|---|
| Alumina (6.5 µm) | 3.9 | irregular | $Al_2O_3$ | 6.5 | 2.3 |
| Alumina (15 µm) | 3.9 | | | 15 | 1.7 |
| Alumina (23 µm) | 4.0 | | | 23 | 1.7 |
| Silica beads (1.7 µm) | 2.2 | spherical | $SiO_2$ | 1.7 | 2.8 |
| Fly ash (4.8 µm) | 2.0 | | | 4.8 | 7.1 |
| Glass beads (18 µm) | 2.5 | | | 18 | 1.4 |

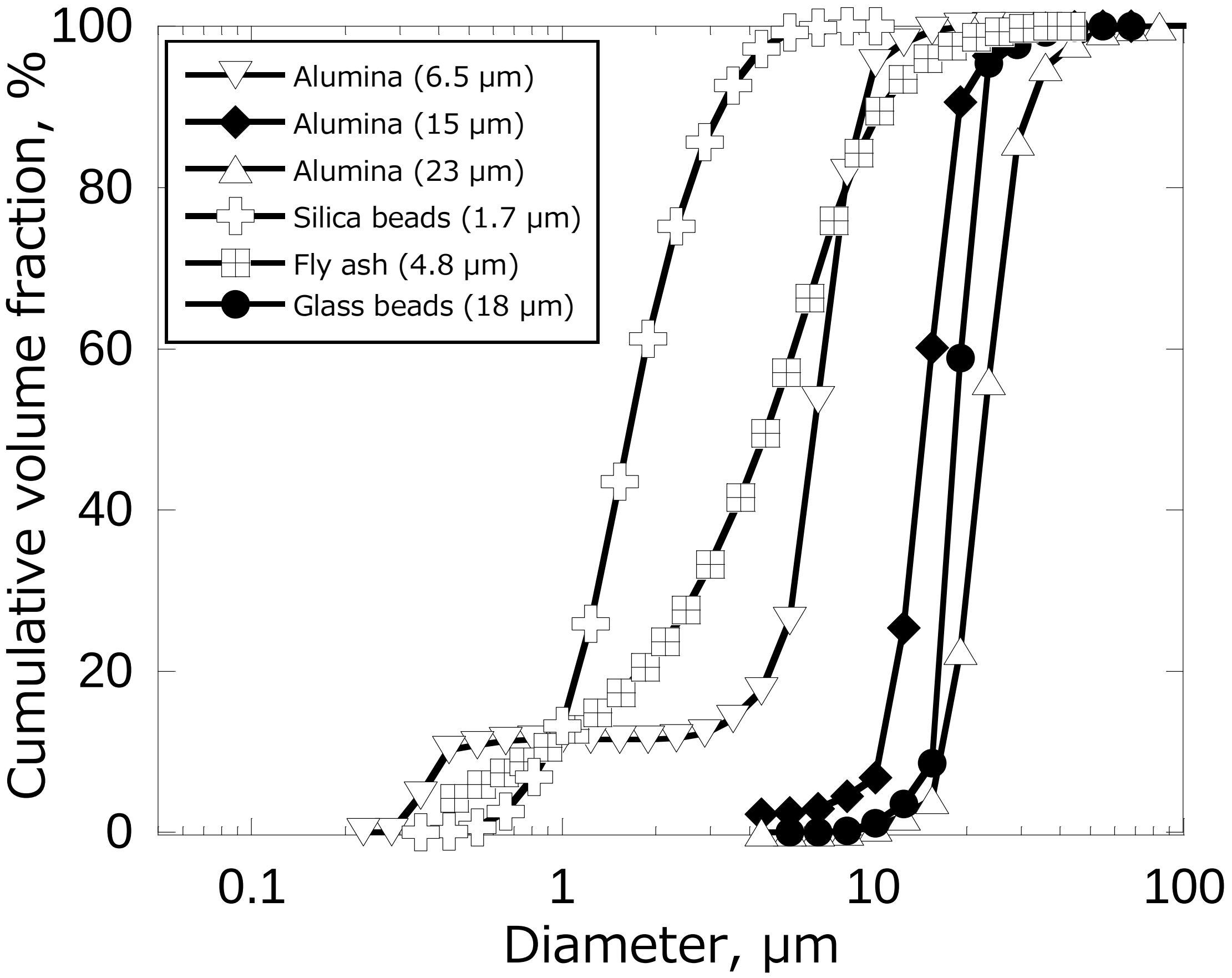

Fig. 2 The particle size distribution of the sample powder.

Because it is easier to obtain commercially available irregularly shaped alumina particles with a narrow size distribution than those consist of $SiO_2$, we used irregular alumina particles with three different median diameters to investigate the size dependence for irregularly shaped particles. Spherical particles were also used in this study for a direct comparison of our experimental results with a theoretical model.

2.2 Sample preparation

We used a stainless cylindrical container with an inner diameter of 19.9 mm and a depth of 17.3 mm (small die) or an inner diameter of 58 mm and a depth of 33 mm (large die) as sample dies. The two dies with different diameter-to-depth ratios allow us to examine the wall effect on compression curves. The sample die used for each sample is indicated in Table 2. We used a new sample powder (opened within 1 week after purchase) or a powder dried at 200 °C over 8 hours before experiments. Sample powders were sieved into a die from ~5 cm height using a sieve with a mesh opening of 500 μm. We heaped the powders on the die and then leveled off the top part of the bed with a spatula so that it does not exceed the height of the die. We adjusted the porosity of samples by tapping because the initial porosity of the samples was not exactly the same each other. We then used piston loading for a finer adjustment of the porosity. The porosities of pre-compression samples calculated by the volume and the mass of the samples using particle density are shown in Table 2.

Table 2 Experimental condition

| Sample powder | Sample die | Initial sample porosity |
|---|---|---|
| Alumina (6.5 μm) | Small | 0.70 |
| | | 0.60 |
| | Large | 0.70 |
| | | 0.60 |
| Alumina (15 μm) | Small | 0.65 |
| | | 0.60 |
| | Large | 0.65 |
| | | 0.60 |
| Alumina (23 μm) | Large | 0.65 |
| | | 0.60 |
| Silica beads (1.7 μm) | Small | 0.80 |
| | | 0.70 |
| Fly ash (4.8 μm) | Large | 0.68 |
| Glass beads (18 μm) | Large | 0.53 |

It is known that compression by piston loading makes a sample stronger due not only to a decrease in the porosity but also to an increase in interparticle forces (Tsubaki and Jinbo, 1984). Samples made of silica particles (1.7 μm) were hard to be compacted to a porosity of 0.7 with tapping. Therefore, it should be noted for this sample that a reduction in the height by piston loading was larger compared with other samples. This may have affected the sample strength.

## 2.3 Experimental procedure

All the experiments were conducted at room temperature in air. The relative humidity was controlled by air conditioners to be less than ~50%.

We conducted our compression experiments using piston loading. The diameter of the piston was slightly smaller than the inner diameter of our sample dies. The loading pressure ranged from 30 to $4\times10^5$ Pa. Sample powders did not leak out of the dies, so that the slightly small piston diameter did not affect the accuracy of sample volume measurements significantly.

We used two different compression apparatuses for low and high pressure regimes. Schematic diagrams of our experimental setups are shown in Fig. 3. We used a light acrylic piston and weights for the low pressure regime. The uniaxial pressure acting on the surface of a sample was 30–3250 Pa in case of the small die and 30-840 Pa in case of the large die. The diameters of these pistons were ~0.3 for the small die and ~1 mm for the large die smaller than the inner diameters of the sample dies. Sample were compressed by a piston and weights. We increased the pressure on the samples by increasing the mass of the weights. The height of the piston was measured using a laser displacement meter. We used a stainless piston fixed to a compressive testing machine for the high pressure regime. The uniaxial pressure on the surface of a sample ranged from 200 or 1000 to $4\times10^5$ Pa. The diameters of these pistons were 0.05 and ~1 mm smaller than the inner diameter of sample dies in case of the small and the large die, respectively. The loading speed was 0.01 mm/sec. The force acting on the sample and the displacement of the piston were recorded. Except for experiments with a compressive testing machine using the large die, we conducted each experiment 3 times using new samples, which were replaced after every compression. For experiments with a compressive testing machine using the large die, we conducted only a single experiment for each sample.

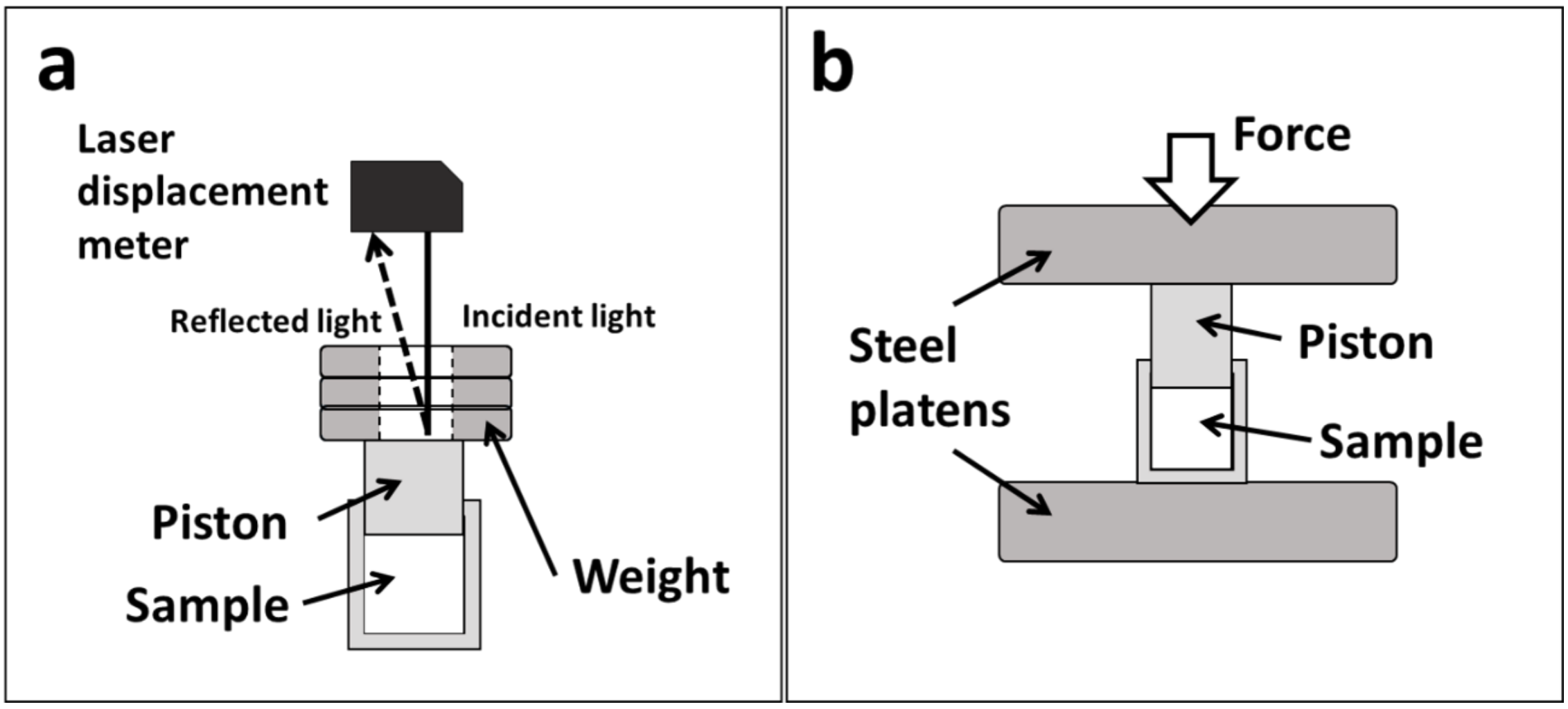

Fig. 3 Schematic diagrams of experimental setups. (a) Compression apparatus for low pressure regime. (b) Setup using compressive testing machine.

We calculated the uniaxial pressure acting on the surface of our sample by dividing the force acting on the piston by the cross-sectional area of the piston. In cases of experiments using weights (low pressure regime), we took into account the vertical pressure distribution in the sample due to its own mass and friction between the sample particles and the wall. We adopted the average of the uniaxial pressure acting on the surface and the bottom of the sample. The pressure acting on the bottom of the sample was calculated using Janssen equation (e.g. Duran, 2000):

$$\sigma_v = \frac{\rho_b g D}{4\mu_w K_a}\left\{1 - \exp(-\frac{4\mu_w K_a}{D}h)\right\} + p_0 \exp(-\frac{4\mu_w K_a}{D}h), \quad (1)$$

where $\rho_b$ is the bulk density of the sample, g is the gravitational acceleration (=9.8 m/s$^2$), D is the diameter of the sample die, $\mu_w$ is the friction coefficient among particles and a side wall, $K_a$ is the ratio of horizontal to vertical stresses, $h$ is the sample height, and $p_0$ is the pressure acting on the surface of the sample. We assumed $K_a$ as 0.5 (an intermediate value of regolith simulant and glass beads) and $\mu_w$ as 0.29 (Sakatani et al., 2016; T. Aoki, private communication). In case of experiments using a compressive testing machine, we referred the uniaxial pressure acting on the surface of the sample. Note that when $p_0 >$ 500 Pa, the difference between the top and bottom pressure increases with $p_0$ and up to ~ 40 % for the case of the small die. The volume of a sample was calculated by the sample height and the cross-sectional area of the sample. The sample height was calculated by the height of the piston or the displacement of the load cell of the compressive testing machine. The porosity at each pressure was estimated from the mass and volume of the sample.

## 3 RESULTS

3.1 Compression behavior of the sample

Results of our experiments are shown in Fig. 4, where symbols are the results obtained using weights and curves are those obtained using a compressive testing machine. The results of the small and the large die are shown in black and orange, respectively. Raw data of the experiments using weights are shown with error bars corresponding to the range of the vertical pressure in the sample. For the measurements using a compressive testing machine, the average value was calculated in case of measurements using the small die by the following way. First, we rounded off the porosity value of each data point to the third decimal place. We chose porosity because the porosity of each data point always monotonically decreased, whereas the pressure did not especially at the low-pressure range. Next, we averaged the pressure values of data points that have the same porosity value. The error bars shown in Fig.4 in gray correspond to the standard deviations of the averaged pressure values. We conducted only a single measurement for each sample using the large die and therefore raw data are plotted. The pressure range of the datasets obtained by weights and a compressive testing machine overlapped from 1000 to ~3000 Pa and from 200 to $\sim 10^3$ Pa for the small and the large die, respectively. The porosity of samples in both measurements were almost the same, but sometimes the effect of friction between the piston and the sample die was seen for the dataset obtained by a compressive testing machine in case of the small die, where the porosity obtained by the machine becomes larger (up to ~3 %) than that obtained by weights at the same pressure. We plot only the results of weights for this pressure range for both experiments because in case of experiments using the small die, the diameter of the piston used for the measurement was smaller than the one used for the compressive testing machine measurements and we expect the data obtained by the piston to be less affected by friction between the piston and the sample die.

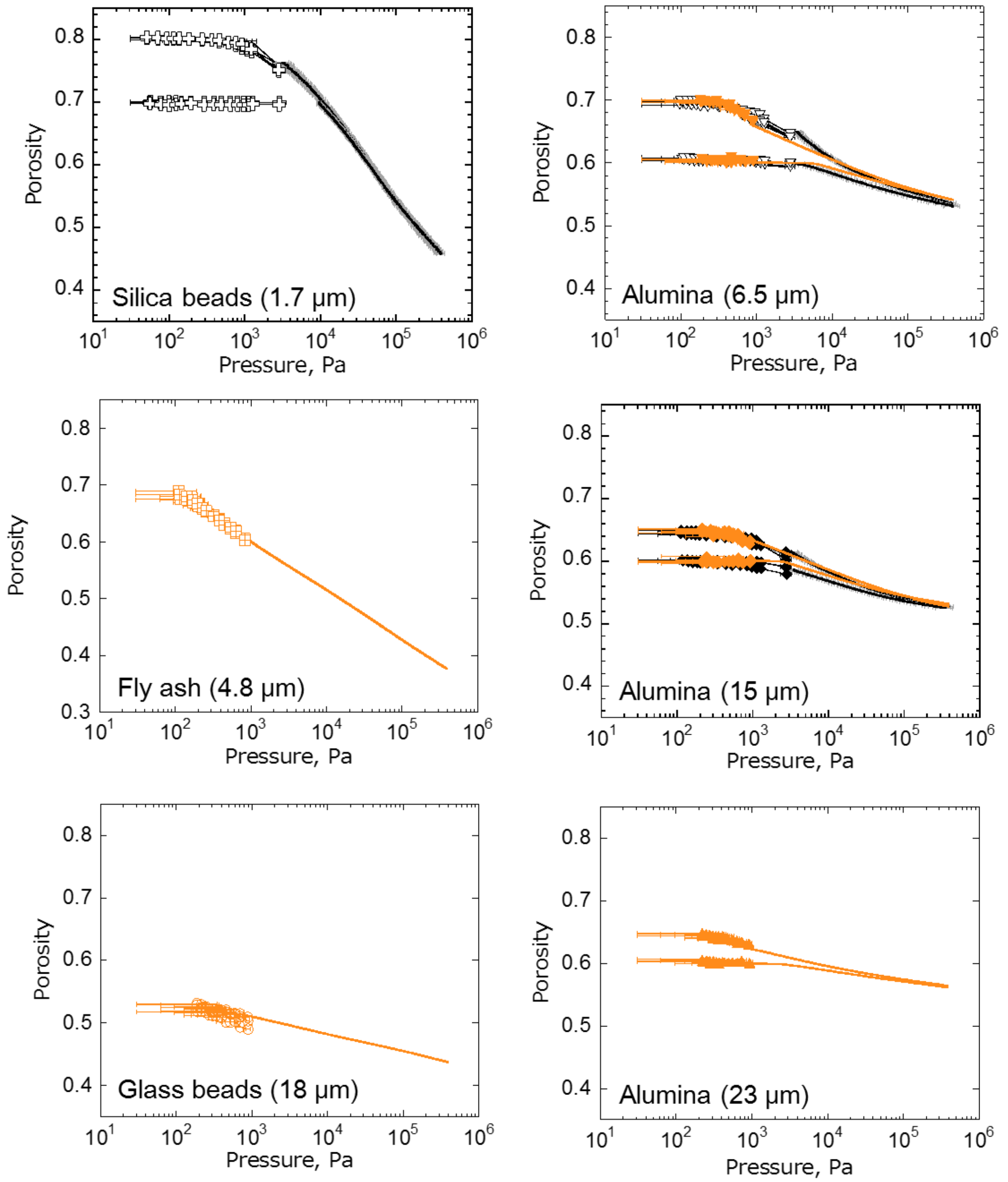

Fig. 4. Compression curves obtained by experiments. Symbols are the results obtained using weights and curves are those obtained using a compressive testing machine. The results of the small and the large die are shown in black and orange, respectively. Samples with different initial porosity and made by the same powders show different compression curve in low pressure regime.

For all samples, the porosity was almost constant at its initial value until the pressure passes the threshold and then decreased above the threshold pressure, at which compaction started. The compression curves of the same powder samples with different initial porosities did not overlap except for a high pressure domain. That is, the compression curve is controlled not only by the properties of constituting particles but also the initial porosity of samples, in other words, the initial configuration of the particles.

In cases of the smallest alumina (6.5 μm) and the medium-sized alumina (15 μm), results of our experiments obtained by the small die (shown in black in Fig. 4) and the large die (shown in orange in Fig. 4) show similar compression curves. The similarity in the compression curves indicates that the wall friction effect due to the measurement system on the results is negligible after we took it into account using Eq. (1).

3.2 Yield strength of each sample

We defined yield strength as the uniaxial pressure on a sample at which the porosity becomes 98% of its initial value, and consider two regimes where the porosity remains largely unchanged and it decreases with pressure. When the yield strength is large enough to be in the pressure regime of the compressive testing machine, we calculated the yield strength as the mean of the uniaxial pressure acting on the sample surface and the bottom of the sample using Eq. (1) in the same way as in the case of the experiment using weights. The yield strength is summarized in Table 3. When data were obtained by both the machine and weights, we chose the one obtained by weights. For the smallest and medium-sized alumina (6.5 μm and 15 μm), only the yield strength obtained by the large die is shown. In cases of alumina (6.5 μm, 15 μm and 23 μm) with an initial porosity of 0.6, the yield strength was obtained only by a single experiment using a compressive testing machine and therefore the uncertainty due only to the pressure distribution in the sample is shown in Table 3. In other cases, we compared this uncertainty with the scatter among repeated experiments and show the larger one in Table 3.

Table 3 Yield strength of each sample.

| Sample powder | Initial porosity | Yield strength (Pa) |
|---|---|---|
| Alumina (6.5 μm) | 0.70 | $(47.0\pm6.4)\times10^{1}$ [a] |
| | 0.60 | $(11.4\pm1.2)\times10^{3}$ [b] |
| Alumina (15 μm) | 0.65 | $(66.9\pm7.2)\times10^{1}$ [c] |
| | 0.60 | $(46.0\pm4.3)\times10^{2}$ [b] |
| Alumina (23 μm) | 0.65 | $(70.2\pm7.4)\times10^{1}$ [c] |
| | 0.60 | $(9.4\pm1.1)\times10^{3}$ [b] |
| Silica beads (1.7 μm) | 0.80 | $(13.2\pm2.6)\times10^{2}$ [c] |
| | 0.70 | $(103.0\pm8.8)\times10^{2}$ [a] |
| Fly ash (4.8 μm) | 0.68 | $(42.5\pm7.0)\times10^{1}$ [a] |
| Glass beads (18 μm) | 0.53 | $(18.3\pm4.2)\times10^{1}$ [a] |

[a] The weighted average value and the uncertainty are shown.
[b] The uncertainty due to pressure distribution in the sample dominates and is shown here.
[c] The discrepancy among the three measurements is larger than the uncertainty of the each measurement due to the pressure distribution in the sample, therefore simple average and the standard deviation of the three values of yield strength is shown.

The yield strength increased as the initial porosity decreased, or the initial filling factor (which is defined as 1-porosity) increased, as far as the same powder is concerned. A higher strength of more compact samples is consistent with an increase in the coordination number according to the filling factor discussed in section 4. The yield strengths of three alumina samples were not significantly different each other when they had the same initial porosity.

## 4. DISCUSSION

### 4.1 Comparison with a model

In this section, first we present the average interparticle force acting on contact points of constituent particles in a sample powder bed under the uniaxial pressure of yield strength using Rumpf's equation (Rumpf, 1970). Next, we compare the obtained interparticle force with a theoretical value and show that the interparticle force between the particles under yield strength can be constrained by the theoretical values.

#### 4.1.1 Average interparticle force derived by the measurement

Rumpf's equation is originally an equation of the tensile strength on the assumption that the stress acting on a cross-sectional surface of random packing of homogeneous spheres is integration of interparticle forces of particles that exist in the surface and given by

$$\sigma_t = \frac{1-\varepsilon}{\pi} N_c \frac{P}{d^2}, \quad (2)$$

where $\sigma_t$ is the tensile strength, $\varepsilon$ is the porosity, $N_c$ is the average coordination number of the powder bed, $P$ is the interparticle force acting on the constituent particles and $d$ is the diameter of the particles. This equation can be also applied to the compression strength in which the force acts on the bed in the opposite direction (Tsubaki, 1984). In this case, the equation is modified as

$$\sigma_c = \frac{1-\varepsilon}{\pi} N_c \frac{F}{d^2}, \quad (3)$$

where $\sigma_c$ is the compression stress acting on a powder bed and $F$ is the force between particles in opposite directions to the compression stress. The average coordination number is related to the porosity and several authors proposed different relationships between the coordination number and the porosity, but the differences are within a factor of three (Suzuki et al., 1980). We used the one proposed by Rumpf (1970) ($N_c=\pi/\varepsilon$) in this paper. The average interparticle force acting on constituent particles under the pressure of yield strength was calculated using Eq. (3). However, there exists inhomogeneity in the distribution of interparticle forces called "force chain", so that Eq. (3) gives just the average of the interparticle forces (Majmudar and Behringer, 2005).

#### 4.1.2. Theoretical force required for rearrangement

A powder bed is compacted by rearrangement of constituent particles, so that the yield strength of the powder bed should be correlated with forces required for the rearrangement of the particles. In the case of compaction, particles should be rearranged by rolling or sliding, depending on the coordination number of the particles. When the coordination number of particles is small and the particles are allowed to roll, the particles move by rolling because the force required for rolling is smaller than that for sliding. However, when the coordination number exceeds 6, the particles move by sliding because the particles are locked in triaxial directions. Here, we consider the rolling friction force ($F_{roll}$) and the sliding friction force ($F_{fric}$) as the forces required to rearrange particles by rolling and sliding, respectively. These forces are given by a theoretical model based on JKR theory with the use of ideal spherical particles having a perfectly smooth surface (Johnson et al., 1971; Dominik and Tielens, 1995; 1996; 1997). This model is well suited for small (~ submicron) spherical particles of diameter which is enough small to ignore surface roughness because the contact radius of the particles is ~50 Å.

The rolling friction force is given by

$F_{roll} = 6\pi\gamma\xi$, (4)

where $\gamma$ is the surface energy of a particle and $\xi$ is the critical rolling displacement. This is a resistance force against rolling due to adhesion between the particles. This resistance acts as a spring when the rolling displacement is smaller than the critical value and therefore enough torque to exceed the criticality is needed to roll a particle. The rolling friction force is a tangential force for rolling required to produce a torque with a moment arm equivalent to the length of particle radius. The critical rolling displacement has not been defined specifically and considered to lie in the range from ~2 Å to a contact radius (Heim et al., 1999). We assumed hereafter the contact radius in equilibrium to be the critical rolling displacement.

The sliding friction force is a force required to slide contacting particles against friction due to atomic scale steps and given by

$F_{fric} = \frac{G^* a^2}{2\pi}$, (5)

where $G^*$ is the reduced shear modulus and $a$ is the contact radius (Dominik and Tielens, 1997). Here contacting particles are made of the same materials so that the reduced shear modulus is $G^*=G/2$ (G: shear modulus of the material) and then the sliding friction force is given by

$F_{fric} = \frac{G a^2}{4\pi}$. (6)

The magnitudes of the rolling and sliding friction forces are dependent on the contact radius $a$ of the particles. In the case of contacting spherical particles consisting of the same material and having the same diameter, the contact radius is given by

$a = \left\{\left(\frac{3r(1-v^2)}{4E}\right)\left[F^* + 3\pi\gamma r + \sqrt{(3\pi\gamma r)^2 + 6\pi\gamma r F^*}\right]\right\}^{\frac{1}{3}}$, (7)

where $r$ is the radius of the particles, $v$ is the Poisson's ratio, $E$ is the Young's modulus, and $F^*$ is an external applied force. The external applied force is given by the product of pressure applied to the sample and the cross sectional area of the particle. In the case of particles under low pressure, this value is negligibly small compared with the other terms in this equation ($3\pi\gamma r \gg F^*$). In this paper, we regard $F^*=0$ to estimate the contact radius.

In the calculation of rolling and sliding friction forces on our sample powder, we assumed spherical particles with a representative diameter of the powder and adopted $\gamma=0.025$ J/m$^2$, $E=73\times10^9$ Pa, and $v=0.17$ for silica powders and $\gamma=0.041$, $E=4.0\times10^{11}$ Pa and $v=0.23$ for alumina powders (Kendall et al., 1987; Spinner, 1962; Burnham et al., 1990; Kamigaito and Kamiya, 1998). The ratios $F_{roll}/F_{fric}$ of our sample powders are of the order of 0.001 except for silica beads (1.7 μm), for which the ratio is of the order of 0.01.

Particles can be rearranged against rolling and sliding friction forces only by the tangential component of a force due to a compressive stress because the rolling and sliding friction exerts a force in a direction parallel to the contact surface of the particles. Therefore, the rearrangement of the particles begins when the component of $F$ parallel to the contact surface exceeds rolling or sliding friction forces. However, hereafter we ignored the effect of contact angle between the particles and assumed the onset of compaction by rearrangement of the particles when the applied force exceeds the rolling or the sliding friction force.

4.1.3 Comparison between averaged interparticle force and theoretical interparticle force

We compared theoretical rolling and sliding friction forces given by Eqs. (4) and (6) for each sample powder with a calculated average interparticle force acting in the direction against compression stress at the yield stress. Figure 5 shows the average interparticle force normalized to the rolling friction force (left) and to the sliding friction force (right). The averaged interparticle forces normalized to $F_{roll}$ exceeded 1 except for fly ash (4.8 μm) and the ones normalized to $F_{fric}$ were always lower than 1. The results suggest that under the stress level of yield strength, the averaged interparticle force due to compression stress is strong enough to rearrange all contact points in a granular layer by rolling and not enough to rearrange the whole particles but some contact points by sliding. At the yield stress, most particles which are allowed to roll are all rearranged by rolling and the other particles which are prohibited to roll are rearranged by sliding so that the yield strength is the sum of the rolling forces and the sliding friction forces. Additionally, an increase in the average interparticle force with decreasing porosity corresponds to decreasing of the number of particles that are allowed to roll because of an increase in the average coordination number of the particles. If the average coordination number is 2 or less, almost all particles are allowed to roll, so that the average interparticle force becomes close to the rolling friction force. However, such a small coordination number is only realized at very high porosities (~0.9) like the proposed value for planetesimals (Kataoka et al., 2013). Therefore, it is natural that the average interparticle force exceeds the rolling friction force in the porosity range of this study. In the case of fly ash (4.8 μm), a considerable amount of particles smaller than the median diameter of the sample powder should have made compaction easier because the force required to rearrange particles decreases with decreasing particle size. This is the reason that the averaged interparticle force normalized to $F_{roll}$ for fly ash is smaller than 1.

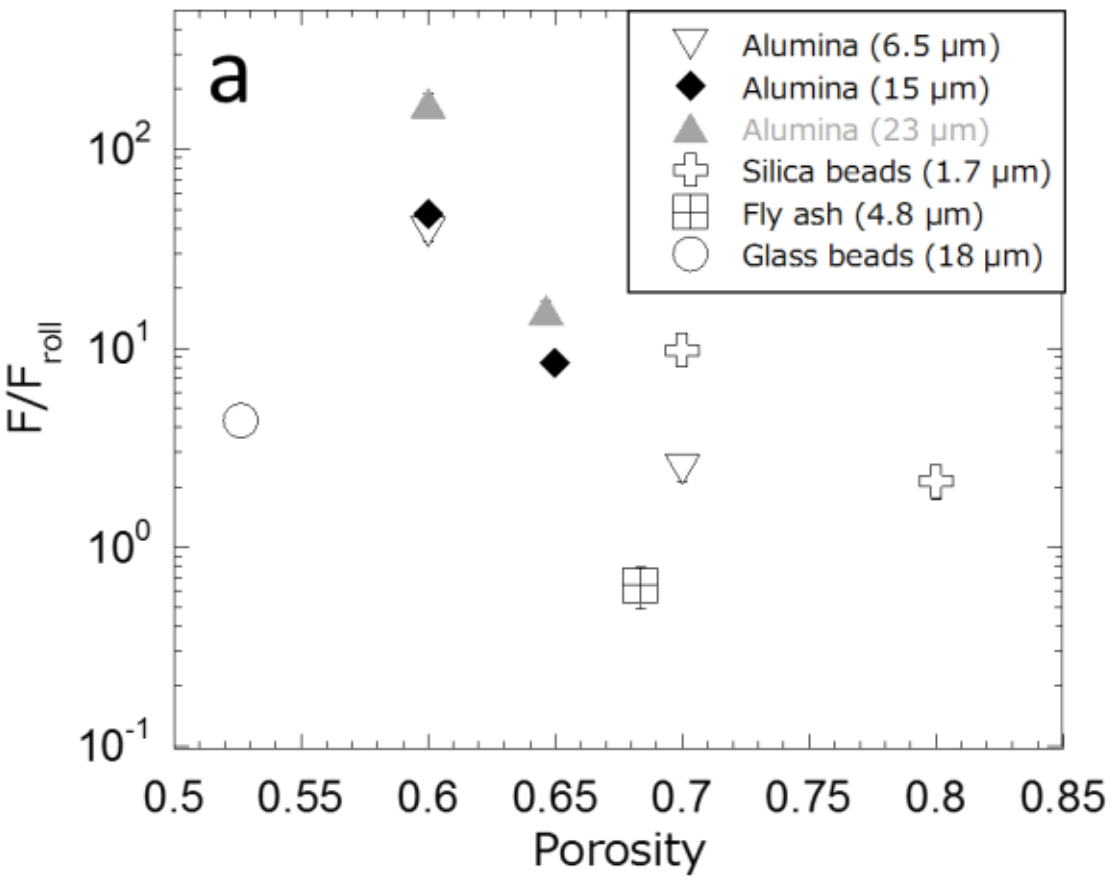

Fig. 5 (a) Average interparticle force acting on contact points calculated by Eq. (3) normalized by the rolling friction force. (b) Average interparticle force normalized by the sliding friction force.

The comparison presented above was too simplified and there are factors to be taken into consideration. Strictly speaking, the theoretical values were valid only for spherical particles, however, alumina particles have irregular shapes, so that the actual force required to rearrange these particles should be different from our estimate. If particles contact at their edges or faces, the contact radius and a force required for rolling or sliding them become smaller or larger, respectively. Additionally, force chains should exist in a granular layer so the force as much as required to slide all contact points are not required for compaction of a granular layer and therefore the ratio $F/F_{fric}$ never reaches 1 even if all particles are prohibited to roll. On the other hand, although we neglected external forces, particles constituting a force chain should have been applied a larger external force than we expected and thus we have underestimated the required force for compaction of these particles.

### 4.2 The pressure-porosity relation after yield

We fit logarithmic functions to the compression curves obtained by a compressive testing machine. The fitted pressure ranged from the smaller one of either the minimum pressure value obtained by a compressive testing machine (840 or 3250 Pa) or the yield strength of each sample to $5\times10^4$ Pa. This pressure range was chosen as an approximation of a logarithmic function holds well. Figure 6 shows the relationship between a slope of compression curves and the sliding friction force. Each symbol is associated with a value of initial porosity and the size of die. The values obtained by our compressive experiment using a small die are shown in black and those using a large die are shown in gray. Only averaged values are shown for those obtained using a small die. The error bars correspond to the standard deviation. The slopes of compression curves tend to decrease as the sliding friction forces of sample powders increase. The compression of a sample becomes more difficult for the samples with a strong sliding friction force. A similar trend was observed in our previous study (Omura et al., 2016). However, we also found that the slopes of compression curves vary with initial porosity of the samples. Moreover, in our previous study, it was shown that samples with a spherical shape and a broad size distribution tended to be compressed easier. A further study is required to separate the effects of particle shape, size distribution, and initial porosity of the sample.

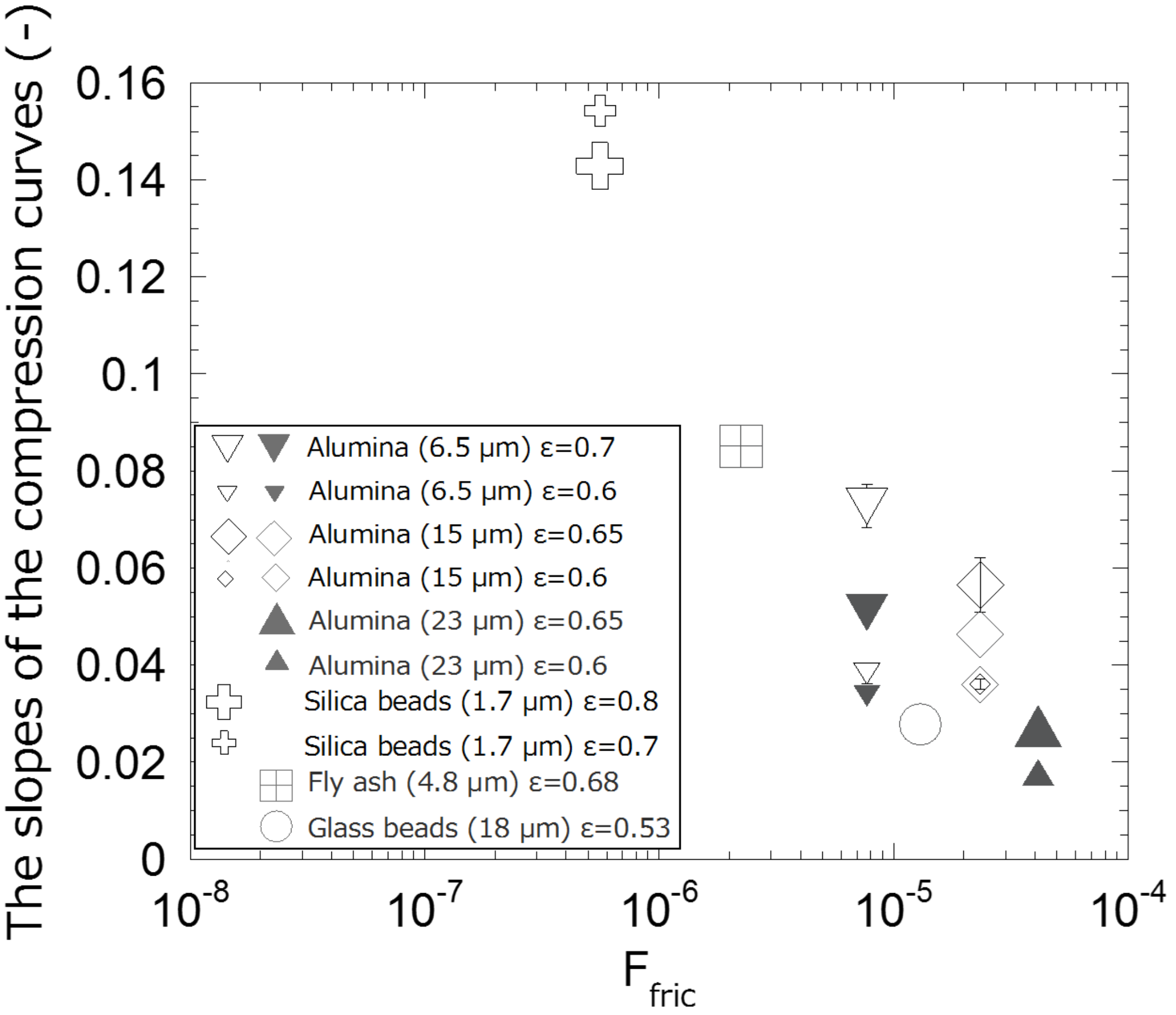

Fig. 6 The relationship between the slope of compression curves after yield and the sliding friction force. Each symbol shows the value corresponding to different initial porosity (ε) with large or small die. Values obtained by compressive experiment using small die are shown in black and using large die are shown in gray.

## 5. APPLICATION TO REGOLITH SURFACE OF SMALL BODIES

Once we better constrain the yield strength of a granular layer, we can estimate the thickness of the layer with a constant value of porosity under gravitational acceleration by comparing the overburden pressure with the yield strength. For instance, if particles having physical properties characteristic of the smallest alumina (6.5 μm) and a porosity of 0.7 cover the surface of a body similar to Phobos and a layer of the particles continued to the depths of the layer, we calculate the thickness of the layer with a constant porosity to be 69±9 m using the gravitational acceleration of Phobos (i.e., $5.8\times10^{-3}$ m/s$^2$), and its density as well as the yield strength obtained by our experiments. If the same granular layer covers Ceres with its gravitational acceleration of $2.7\times10^{-1}$ m/s$^2$ based on its mass and mean radius, the thickness of the granular layer with constant porosity becomes thinner and is estimated to 1.50±0.2 m (Thomas et al., 2005). The upper and lower limits of yield strength of a granular layer with a given porosity and particle diameter may be constrained by theoretically estimated rolling and sliding friction forces (Eqs. 4 and 6) and using Eq. (3). The yield strength of a granular layer approaches the upper limit as the porosity of the layer decreases. According to our results obtained by alumina samples, a granular layer having a porosity of 0.6 should have a yield strength of approximately one tenth of the value estimated by sliding friction force. Such an estimation allows us to constrain the thickness of a regolith layer with constant porosity.

## 6. CONCLUSIONS

Compression experiments were conducted using powders with different size distributions, grain shapes (spherical or irregular), and material composition. We adjusted the initial porosity of samples by tapping and piston loading in the range from 0.55 to 0.85. For some sample powder, we obtained compression curves starting from different initial porosities. The applied uniaxial pressure ranged from 30 to $4\times10^5$ Pa.

In this study, our focus was put on the low-pressure and high-porosity regime of the compression curves. The porosity of the samples remained almost unchanged until the applied pressure reaches the threshold. We defined yield strength as the uniaxial pressure on a sample with 98% of its initial porosity for separating two regimes where the porosity is kept constant or decreases with pressure. The yield strength increased as the initial porosity decreased as far as the same material is concerned. The yield strengths of three irregular alumina samples were not significantly sensitive to their sizes when they had the same initial porosity.

We calculated the average interparticle forces acting on contact points of constituent particles of powder beds under the pressure of yield strength and compared them with a previously proposed theoretical model of forces required to rearrange spherical particles. The average interparticle force increased with decreasing porosity and was larger than the estimated rolling friction force and smaller than the estimated sliding friction force of constituent particles. Accordingly, the yield strength of a granular bed consisting of certain particles may be constrained by rolling and sliding friction forces of constituent particles. This constraint may be useful for planetary scientists to estimate the compressive property of asteroidal regolith and the depth at which the porosity is almost equal to that of the asteroid surface, and to interpret the compressive properties of an asteroid surface probed by landing instruments such as MASCOT (Ho et al., 2016).

## ACKNOWLEDGMENTS

We are grateful to M. Hyodo for allowing us access to the laser diffractometer and scanning electron microscope. We thank to H. Kimura for valuable comments on this paper. This research was supported by the Hosokawa Powder Technology Foundation, the Space Plasma Laboratory, ISAS, JAXA, Japan, and a grant-in-aid for scientific research from the Japanese Society for the Promotion of Science of the Japanese Ministry of Education, Culture, Sports, Science, and Technology (MEXT) (No. 25400453).